\documentclass[11pt]{article}
\usepackage{epsfig,rotating}

\def\gapproxeq{\lower .7ex\hbox{$\;\stackrel{\textstyle >}{\sim}\;$}}
\def\lapproxeq{\lower .7ex\hbox{$\;\stackrel{\textstyle <}{\sim}\;$}}

\begin{document}
{\noindent \large \hspace*{13.5cm} BA-00-37 \newline \hspace*{13.5cm} DESY 00-125}
\begin{center}
{\Large \bf  Tracing Tau-Neutrinos from WIMP-Annihilation} 
 \end{center}
\vspace{0.5cm}
\begin{center}
{Marek Kowalski}
\end{center}
\renewcommand{\thefootnote}{\fnsymbol{footnote}}
\begin{center}
{\it DESY Zeuthen\footnote{Permanent address}

Platanenallee 6, D-15738 Zeuthen, Germany 

E-mail: marek.kowalski@desy.de

and Bartol Research Institute

University of Delaware, Newark, DE 19716}
\end{center}
\renewcommand{\thefootnote}{\arabic{footnote}}
\begin{center}
\end{center}
\begin{abstract}
\noindent
Accumulation and annihilation of weakly interacting massive particles (WIMPs) in the earth and the sun may be observed by the resulting neutrino signal.
We demonstrate that, for certain parts of the SUSY parameter space, present and future neutrino telescopes can expect a higher rate of events due to tau-neutrinos than of events due to muon-neutrinos. We show how $\nu_{\tau}\leftrightarrow\nu_{\mu}$ oscillations, and in the case of the sun also scattering, absorption and regeneration, modify the signal expectation.
We find that most currently proposed neutralino models  predict equal or increased muon fluxes if $\nu_{\tau}\leftrightarrow\nu_{\mu}$ oscillations are included.
\end{abstract}
\section{Introduction}
Non-baryonic dark matter is believed to exist in order to explain the available
 data on large scale structures in the universe \cite{costri}.
A good candidate for such  non-baryonic dark matter is  the lightest supersymmetric particle (LSP) and in particular the lightest neutralino, $\tilde{\chi}_{0}$.
Cosmological arguments restrict the mass of the neutralino to be lighter than a few TeV \cite{Edsjo:1997bg} and combined with data from accelerators set lower limits  of about 50~GeV \cite{ellis_susy}.
WIMPs populating the galactic halo would loose energy by scattering in the sun and the earth and if slow enough they would be trapped by the gravitational potential. These trapped WIMPs then annihilate with each other to produce quarks, leptons, gauge or higgs bosons which then  decay and thereby  produce neutrinos.   
The differential energy flux  of these neutrinos is given by
\begin{equation}
\frac{dN_{i}}{dE}=\frac{\Gamma}{4\pi R^{2}}\sum_{F}B_{F}\frac{dN_{F,i}}{dE}
\label{eq:rate}
\end{equation}
where $\Gamma$ is the annihilation rate within the sun or earth, $R$ is the distance to the detector location, $B_{F}$ is the branching ratio for WIMP annihilation into the different final states $F$ and $\frac{dN_{F,i}}{dE}$ is the differential number of neutrinos of flavor $i$ produced in the decay of $F$.
Detailed calculations of observable event rates exist for a wide range of parameters of the Minimal Supersymmetric Standard  Model (MSSM) \cite{edsjo,Bottino}. The evaluated signal typically consists of neutrino-induced muons which, due to their long range, have an experimentally  promising signature.

Here we want to draw attention to the tau-neutrino for the following reason.
Annihilation of WIMPs can be a significant source of tau-neutrinos resulting in
a flux of tau-neutrinos which is up to 6 times higher than the flux of muon- or electron-neutrinos. This is realized for those parts of the allowed MSSM parameter space in which the branching ratio for annihilation into a pair of taus is large. The tau decay yields always a tau-neutrino and in 17\% also an electron- or muon-neutrino. 
A high tau-neutrino flux has strong implications since present data of atmospheric neutrinos provide evidence for strong $\nu_{\tau}\leftrightarrow\nu_{\mu}$ mixing \cite{osci_par}. Tau-neutrinos would oscillate to muon-neutrinos and contribute to the observable muon-signal. Calculations of limits on MSSM parameters from observed muon fluxes  \cite{Ambrosio:1999qj,Bai:2000mb,Okada:2000ve} have to take this effect into account. 
In addition, present and future high-energy neutrino telescopes as AMANDA \cite{AMANDA}, ANTARES \cite{ANTARES}, BAIKAL \cite{BAIKAL} and NESTOR \cite{NESTOR} have large detecting volumes which make a search for cascades induced by electron- or tau-neutrinos promising. 
It should be mentioned that event rates due to tau-neutrinos have been previously calculated in \cite{Seckel} for a specific SUSY model.

In the next section we discuss some aspects of the MSSM phenomenology relevant for the  production of tau-neutrinos. In section \ref{sec:prop} we discuss the propagation of tau-neutrinos through the interior of the sun, taking into account scattering, absorption and regeneration as well as the possibility of neutrino mixing. We then show how $\nu_{\tau}\leftrightarrow\nu_{\mu}$ oscillations modify the expected muon flux from the sun and the earth. We examine the prospects of direct detection of tau-neutrinos. Following~\cite{Jung_report,Jung_modelind} we choose model-independent, extreme cases of the branching ratios, as well as specific intermediate branching ratios motivated by the MSSM phenomenology.
\noindent
\section{MSSM phenomenology}
\label{sec:MSSM}
\begin{figure}[htbp]
\begin{center}
\centerline{\epsfig{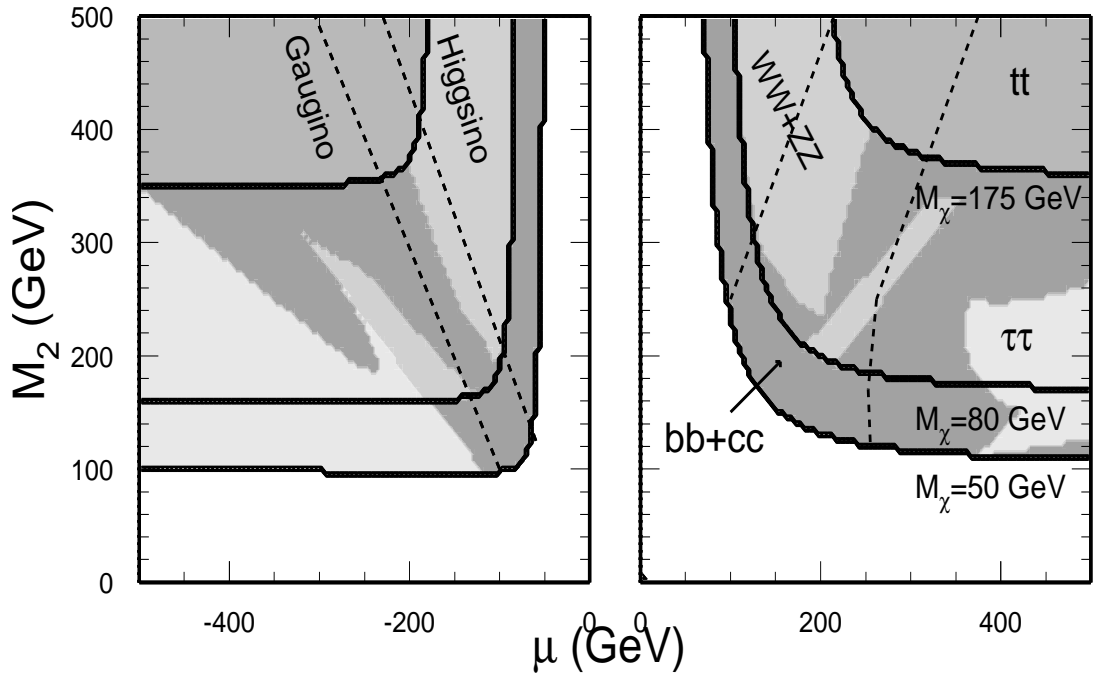}}
\caption{Distribution of branching ratios for the final states  $c\overline{c}$, $b\overline{b} ,t\overline{t},\tau\overline{\tau},W^{+}W^{-}$ and $Z^{0}Z^{0}$. The final states with the largest branching ratio are indicated by different shading. The cross-sections \cite{cross_susy} were evaluated in the limit $v_{\chi} \rightarrow 0$. We assumed $M_{A}$=1000~GeV, $\tan\beta$=3, degenerate sfermion masses of 200~GeV and $M_{1}=\frac{5}{3} M_{2}\times \tan^{2}\theta_{W}$. Contours of equal M$_{\chi}$ are indicated by the full lines. Dashed lines distinguish regions of different gaugino/higgsino character.}
\label{fig:m2mu}
\end{center}
\end{figure}

The four neutralinos are a linear combination of the super-partners of the electroweak gauge bosons and the Higgs boson.
In the basis $(\tilde{B},\tilde{W}_{3},\tilde{h}_{1}^{0},\tilde{h}_{2}^{0})$, the neutralino mass matrix is given by
\begin{equation}
\left(\begin{array}{cccc}
M_1 & 0 &-m_Z\sin\theta_W\cos\beta &m_Z\sin\theta_W\sin\beta \\
0 & M_2 &m_Z\cos\theta_W\cos\beta &-m_Z\cos\theta_W\sin\beta \\
m_Z\sin\theta_W\cos\beta &-m_Z\cos\theta_W\cos\beta&0&-\mu\\ 
m_Z\sin\theta_W\sin\beta &-m_Z\cos\theta_W\sin\beta& -\mu &0  
\end{array}\right),
\label{eq:mass}
\end{equation}
where $M_1$ and $M_2$ are the gaugino mass parameters, $\mu$ is the higgsino mass parameter and tan$\beta$ is the ratio of the Higgs vacuum expectation values.
Here we use the convention and notation of \cite{cross_susy}. 
The eigenvectors of (\ref{eq:mass}) then represent the four neutralinos of which the lightest, $\tilde{\chi}_{0}$, is 
\begin{equation}
\tilde{\chi}_{0}=N_{01}\tilde{B}+N_{02}\tilde{W}_{3}+N_{03}\tilde{h}_{1}^{0}+N_{03}\tilde{h}_{2}^{0}.
\end{equation}
We define the gaugino fraction to be $R_{\chi}=N_{01}^2+N_{02}^2$ and refer to  a neutralino as  gaugino-like if $R_{\chi}>0.9$, as  higgsino-like if $R_{\chi}<0.1$, and as mixed if the gaugino fraction is inbetween 0.1 and 0.9. 
To obtain the distribution of branching ratios we have evaluated the cross-sections given in \cite{cross_susy} for annihilation into $c\overline{c},$ $b\overline{b}$, $t\overline{t},\tau\overline{\tau}$, $W^{+}W^{-}$ and $Z^{0}Z^{0}$ in the approximation of WIMPs being at rest.
Figure \ref{fig:m2mu} shows, by means of different shading, the dominant branching ratios in the $M_2 - \mu$ plane, with the trilinear mass chosen to be $M_{A}$=1000~GeV, $\tan\beta$=3 and degenerate sfermion masses of 200~GeV. The GUT relation $M_{1}=\frac{5}{3} M_{2}\times \tan^{2}\theta_{W}$ was assumed to hold.
The mass contours  of the lightest neutralino, $\tilde{\chi}_{0}$, as well as the gaugino fraction ($R_\chi$=0.1, 0.9) are indicated by the full and dashed lines, respectively.
   
A higgsino-like neutralino is already ruled out provided its relic density lies within the cosmologically interesting region $0.1\le\ \Omega_{\chi}h^{2} \le 0.3$~\cite{ellis_susy}.

To estimate the order of magnitude of the different branching ratios, we evaluate the cross-sections for a few simple neutralino compositions. In  particular we focus on the ratio $B_{b\overline{b}}/B_{\tau\overline{\tau}}=\sigma_{b\overline{b}}/\sigma_{\tau\overline{\tau}}$, since, if the neutralino is gaugino-like and lighter than the top-quark, annihilation into $b\overline{b}$ and $\tau\overline{\tau}$ is responsible for most of the expected neutrino-signal. The final states $b\overline{b}$/$\tau\overline{\tau}$  also produce extremely  low/high fluxes of tau-neutrinos compared to the fluxes of muon-neutrinos. 

Generally, if annihilation happens via s-channel exchange of the $Z^0$ and neutral Higgs bosons, $A$, and if mass threshold effects are neglected, the ratio is given by  
\begin{equation}
\frac{B_{\tau\overline{\tau}}}{B_{b\overline{b}}}\simeq\frac{1}{c_f}\frac{m_\tau^2}{m_b^2},
\label{eq:higgs}
\end{equation}
 where $c_f=3$ is the number of colors and $m_\tau$ and $m_b$  are the tau and b-quark masses, respectively. The mass dependency is a result of the helicity-suppressed annihilation into two fermions. Here, QCD correction to $\sigma_{b\overline{b}}$ becomes significant but can be absorbed into a ``running mass'' \cite{cross_susy}.  To give an example, for a neutralino of 80~GeV, $m_b$ reduces to 3/4 of its tree-level value. We thus obtain a ratio of about $B_{\tau\overline{\tau}}/B_{b\overline{b}} \simeq 0.1$ .

If the neutralino would be a pure B-ino ($N_{01}=1$ and $N_{02}=N_{03}=N_{04}=0)$, the charge dependent t- and u-channel exchange of sfermions dominates
 the cross section, and we obtain 
\begin{equation}
\frac{B_{\tau\overline{\tau}}}{B_{b\overline{b}}}\simeq\frac{1}{c_f}\frac{m_\tau^2m^4_{\tilde{q}}}{m_b^2m^4_{\tilde{l}}}\left(\frac{(T_3-e_\tau)^2+e_\tau^2}{(T_3-e_b)^2+e_b^2}\right)^2.
\label{eq:bino}
\end{equation}
Here, $T_3$ is the weak isospin ($=-1/2$) and $e$ is the electrical charge ($e_\tau=-1$ and $e_b=-1/3$). It was assumed that the slepton and squark masses, $m_{\tilde{l}}$ and  $m_{\tilde{q}}$,  are horizontally degenerated. For  $m_{\tilde{l}}=m_{\tilde{q}}$ and $m_{\chi}=$80~GeV it follows that $B_{\tau\overline{\tau}}/B_{b\overline{b}} \simeq 9$.
It should be noted  that if the GUT  relation $M_{1}=\frac{5}{3} M_{2}\times \tan^{2}\theta_{W}$ holds and the  neutralino is gaugino-like it usually also resembles the B-ino, but with a small higgsino contribution remaining. This higgsino contribution typically weakens relation (\ref{eq:bino}). By scanning over a large range of MSSM parameters we find a more characteristic ratio for B-ino like neutralinos of $B_{\tau\overline{\tau}}/B_{b\overline{b}} = \mathcal{O}(1)$.
Note also, that this ratio increases for $m_{\tilde{l}}<m_{\tilde{q}}$ as predicted in mSUGRA models \cite{Drees:1995hj}.

If the neutralino is W-ino like ($N_{02}=1$ and $N_{01}=N_{03}=N_{04}=0$) and  for a neutralino mass range $m_{W,Z}<m_\chi<m_{t}$, annihilation into gauge bosons generally dominates. For $m_\chi<m_{W,Z}$, the annihilation into fermions is characterized by equation (\ref{eq:higgs}).
It should be noted that even though the GUT relation $M_{1}=\frac{5}{3} M_{2}\times \tan^{2}\theta_{W}$ prohibits the lightest neutralino to be W-ino like \cite{cross_susy}, there are models with anomaly-mediated supersymmetry breaking which predict a W-ino like LSP \cite{Gherghetta:1999sw}. 

If the neutralino is of mixed composition, we can expect a large branching ratio for annihilation into $W^+W^-$ and $Z^0Z^0$ \cite{Feng:2000zu}. 
Finally, if the neutralino is heavier than the top mass, the branching ratios for annihilation into a pair of top quarks is almost always favored.   
 
Summarizing, it can be said that the branching ratio for $\tau\overline{\tau}$ is highest when the lightest neutralino is primarily B-ino and not heavier than the top-quark. Thus it is this area of the MSSM parameter-space for which significantly higher fluxes of tau-neutrinos can be expected. 

\noindent
\section{Neutrino propagation: scattering, absorption, regeneration and oscillation}
\label{sec:prop}
Neutrinos in the energy range relevant for WIMP-annihilation propagate nearly unscattered through the earth. However, for the sun, its large size and high density in the core make it necessary to include neutrino reactions \cite{Seckel}. 
The  main differences between the reactions of tau- and electron- or muon-neutrinos is that the charged current (CC) cross-section for tau-neutrinos is reduced  due to kinematic reasons, and that a CC tau-neutrino reaction will regenerate a tau-neutrino in the prompt decay of the fully polarized tau. Thus the tau-neutrinos have to be treated differently. A particular interesting situation arises when neutrino mixing is considered. The general case of neutrino 
mixing in the presence of media was discussed in~\cite{stodolsky} and leads in case of forward-scattering to the well studied MSW-effect~\cite{MSW}. 
The propagation of oscillating neutrinos can be described by 
their density matrix~\cite{stodolsky}, $\rho_{p}$:
\begin{equation}
\dot{\rho_{p}}=-i[\Omega_{p},\rho_{p}]+\int dp' (W_{p'p}\rho_{p'}-W_{pp'}\rho_{p})
\label{eq:prop}
\end{equation}
\[\hspace{1.7cm}-\sum_{i=1}^{n_{flavor}}\frac{1}{2}\{\rho_{p},I_{i}\}A_{p}^{i}+\int dp' I_{\tau}\rho_{p'}I_{\tau}R_{p'p}\]
where $\Omega_{p}$ is the momentum-dependent matrix of oscillation frequencies, $W_{pp'}$ is the transition probability for NC-scattering of a neutrino of momentum $p$ into a state $p'$, which is a product of the density and the composition-dependent cross-section.  $I_{i}$ is the projection operator onto the flavor $i$, $A_{p}^{i}$ is the absorption rate of flavor $i$ due to a CC-reaction and $R_{p'p}$ is the regeneration probability of a tau-neutrino. 

The first term in equation (\ref{eq:prop}) is responsible for flavor-oscillations while the second term, since the cross-sections for neutral current (NC) reactions are to a good approximation flavor-independent\footnote{We therefore neglect the MSW-effect related case of ${\nu}_{e}-e$ forward scattering.},
 describes coherent scattering. It is similar to a gain-loss term in a usual Boltzmann-type collision integral. The third term accounts for absorption, which is flavor dependent due to the smaller CC cross-section of the tau-neutrino.
 However, the fourth and last term breaks the coherent development by adding pure tau-neutrinos due to the regeneration-mechanism described above. We thereby neglect the muon- or electron-neutrinos which are produced in 17~\% of the tau-neutrino regeneration.

Here, we will restrict ourself to two flavor $\nu_{\tau}\leftrightarrow\nu_{\mu}$ oscillations. For the energies under consideration, this can be justified as mixing of $\nu_{e}\leftrightarrow\nu_{\mu}$ is suppressed due to the low mass differences of $\sim 10^{-5}$~eV$^{2}$ derived from solar neutrino data \cite{osci_par} and due to the relatively short baseline of the high density core of the sun.
Mixing of $\nu_{e}\leftrightarrow \nu_{\tau}$ is also partially suppressed due to constraints on the effective mixing-angle, $\sin^{2}2\theta_{e\tau}\leq 0.1$ \cite{Bilenky}.

Because of the strongly inhomogeneous density and composition of the sun, the analytic treatment of (\ref{eq:prop}) becomes very involved and we thus treat the problem by means of a Monte Carlo calculation. 
We simulate the neutrino propagation by stochastic interactions along its way out of the sun.  
The cross-sections for NC- and CC-reactions were calculated using the MRS parton distribution function~\cite{mrs}. 
The density and composition of the sun as a function of the radius were taken from~\cite{Bahcall_sun}.
The regeneration of the tau-neutrino by tau-decay was simulated according to \cite{Dutta:2000jv}.
\begin{figure}[htbp]
\begin{center}
\centerline{\epsfig{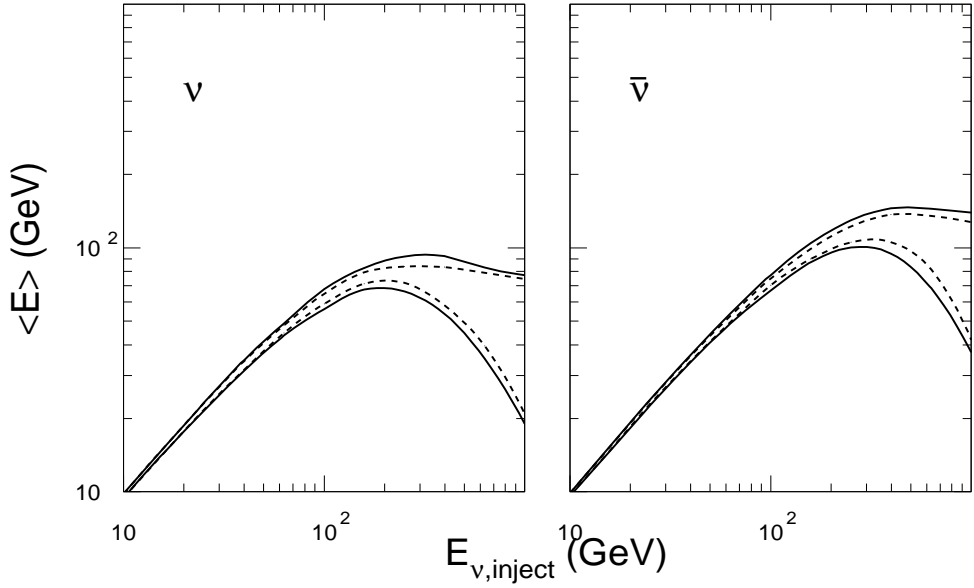}}
\caption{Average neutrino-energy after propagating from the core to the surface of the sun, as a function of the injection energy. The injected neutrino beam is initially single flavored but after propagation represents the sum of the oscillating flavors. The full lines show the case of no oscillation while the dashed lines corresponds to full mixing and $\Delta m^{2}=0.003~$eV$^{2}$.
The upper two curves correspond to an initial tau-neutrino beam and the lower two represent an initial muon-neutrino beam. Electron-neutrino attenuation is similar to the case of muon-neutrino attenuation without oscillation.  Absorbed neutrinos are included in the average with $E_{\nu}$=0.}
\label{fig:prop}
\end{center}
\end{figure}
The dynamics of equation (\ref{eq:prop}) is incorporated in the 
following way: A coherent development of the oscillating neutrino-beam is assumed even when the neutrino undergoes a NC-reaction. 
In case of a CC-reaction the neutrino is absorbed and in the case of a tau-neutrino regenerated as a pure tau-neutrino with a lower energy.

Figure \ref{fig:prop} shows the average neutrino-energy after propagating from the core to the surface of the sun as a function of the injection energy.
It can be seen that, due to tau-neutrino regeneration, the tau-neutrino beam is considerably less attenuated than the muon- or electron-neutrino beam and that the difference between the mixing and no-mixing case is relatively small. 
The latter can be understood by comparing the typical absorption length, $\lambda_{absorption}$,  to the oscillation length, $\lambda_{oscillation}$,  and the size of the scattering region  $d_{scatter}$. Neutrino oscillations
can have a strong influence if  
\begin{equation}
\lambda_{oscillation} \lapproxeq \lambda_{absorption} \lapproxeq d_{scatter}.
\end{equation} 
For a neutrino-energy of 200~GeV, a total cross-section of $0.5\times10^{-38}~$cm$^{2}\times E_{\nu}($GeV$)$, an average core density of 100~g/cm$^{3}$ and a scattering region  of $d_{scatter}\simeq 0.1 \times R_{\odot}$ (which contains approximately 60\% of the total column depth) we obtain
\begin{equation}\lambda_{oscillation} \simeq \lambda_{absorbtion} \simeq 2\times d_{scatter}
\end{equation} 
Hence, it can be argued that above 200~GeV the
influence of neutrino oscillations is reduced due to the long oscillation-length
and  below this energy, scattering becomes less important. This means that in a first approximation, propagation through the sun decouples from neutrino-oscillations and can therefore be treated separately.
We note that a similar line of arguments applies to the influence of oscillation in the scenario of 4 neutrino flavors in which the fourth neutrino is a sterile neutrino.

We conclude this section with a comment on the relevance of the MSW-effect~\cite{MSW}, which has been discussed for
 the relevant energy range in~\cite{ellis_msw}. 
 Resonant flavor-conversion between the electron- and muon-neutrino is generally suppressed due to the high neutrino-energies considered here. 
However, present constrains on the oscillation-parameter space allow resonant flavor conversion between the electron- and tau-neutrino if
 $E_{\nu_{e,\tau}}\lapproxeq$75~GeV~\cite{ellis_msw}.
The conditions for the MSW-effect are met, if at all, in the outer layers of the sun, where the neutrinos practically not suffer from further attenuation. The above results are therefore unaffected, and we will neglect the MSW-effect in the further discussion.

\section{Event Rates}

To calculate event rates for a given detector, formula (\ref{eq:rate}) has to be folded with the detection probability of the neutrino. Here we will assume
a detector of 1~km $\times$ 1~km $\times$ 1~km dimensions and, somewhat arbitrarily,
set its energy threshold to $E_{\nu}=25$~GeV. For a discussion of the effect of the energy threshold see~\cite{edsjo}.  Since detailed muon rate calculations exist for non-oscillating muon-neutrinos~\cite{edsjo,Bottino} we will restrict ourself to the analysis of changes of the event rates due to oscillation or contained tau-events.   

We follow~\cite{Jung_report,Jung_modelind} and evaluate the extreme cases of branching ratios, $B_{F}$, for which signal expectations are either lowest or highest. 
If the neutralino is lighter than the top-quark, the hardest neutrino spectrum can be expected when the neutralino  annihilates into pairs of taus, $W^{\pm}$s or $Z^{0}$s~\cite{Kamion}. 
We thus assume an upper limit given by an exclusive annihilation into taus.
A lower limit is given by annihilation into a pair of $b$-quarks, with the branching ratio being almost always significantly larger than for annihilation into $c$-quarks. 
If the neutralino is heavier than the top-quark the dominant annihilation channel will be the top quark or, if the neutralino is higgsino-like, it might also annihilate into pairs of $W^{\pm}$s or $Z^{0}$s.
We neglect final states which include gluons, since they produce generally a very soft neutrino spectrum which is not suitable for detection in large scale neutrino-detectors~\cite{drees_gl}. We will also omit final states containing Higgs bosons. 
The decay of the neutral Higgs is characterized by an equation similar to (\ref{eq:higgs}) or, if heavy enough, by annihilation into $W^{\pm}$s or $Z^{0}$s. Annihilation into charged Higgs bosons can lead to an increase of the tau-neutrino flux, as was recently discussed in \cite{Bednyakov:2001uw}.   

The decay of heavy-quarks, gauge-bosons and taus was simulated with JETSET~\cite{jetset}. Stopping of the $b$- and $c$-quark in the sun
can become significant and we included it by sampling from a stopping spectrum
developed in ~\cite{Seckel}. The distribution of WIMPs is assumed to be concentrated at the core of the sun which is a valid approximation for the WIMP-masses considered here~\cite{Seckel}.
The neutrino propagation through the sun has been discussed above.
The muon range in water is approximately calculated by 
\begin{equation}L(E_{\mu})=\frac{1}{\beta\rho}\ln\left(\frac{E_{\mu}+\alpha/\beta}{\alpha/\beta}\right)
\end{equation}
with $\alpha$=0.0025~GeVg$^{-1}$cm$^{2}$, $\beta=4.0\times10^{-6}$g$^{-1}$cm$^{2}$ and $\rho$ being the density of the medium~\cite{PDG}.


\begin{figure}[htbp]
\begin{center}
\centerline{\epsfig{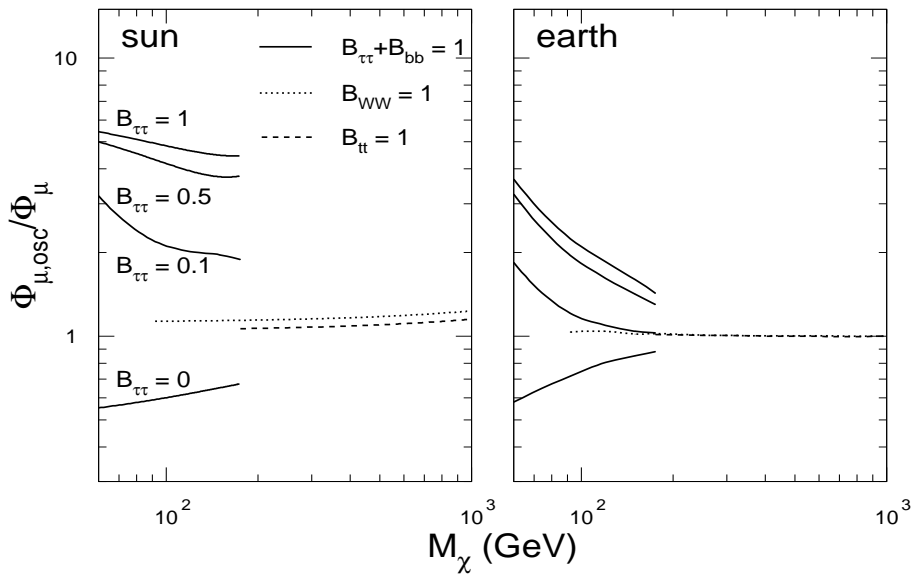}}
\caption{Ratios of neutrino-induced muon fluxes with and without oscillations, $\phi_{\mu,osci}/\phi_{\mu}$, are shown for the sun (left) and the earth (right). The cases of extreme and intermediate branching ratios (assuming $B_{bb}+B_{\tau\tau}=1$) are shown by the full four  
lines, each one for a different $B_{\tau\tau}$. Annihilation into W-bosons (top quarks) is represented by the dotted (dashed) line. Annihilation into Z-bosons leads to a similar ratio as annihilation into W-bosons. 
In the case of $\nu_{\mu}\leftrightarrow\nu_{\tau}$ oscillations, full mixing and $\Delta m^{2}=0.003~$eV$^{2}$ has been assumed. The detector was assumed to have an energy threshold of $E_{\nu}=25$~GeV.}
\label{fig:rate}
\end{center}
\end{figure}

\begin{figure}[htbp]
\begin{center}
\centerline{\epsfig{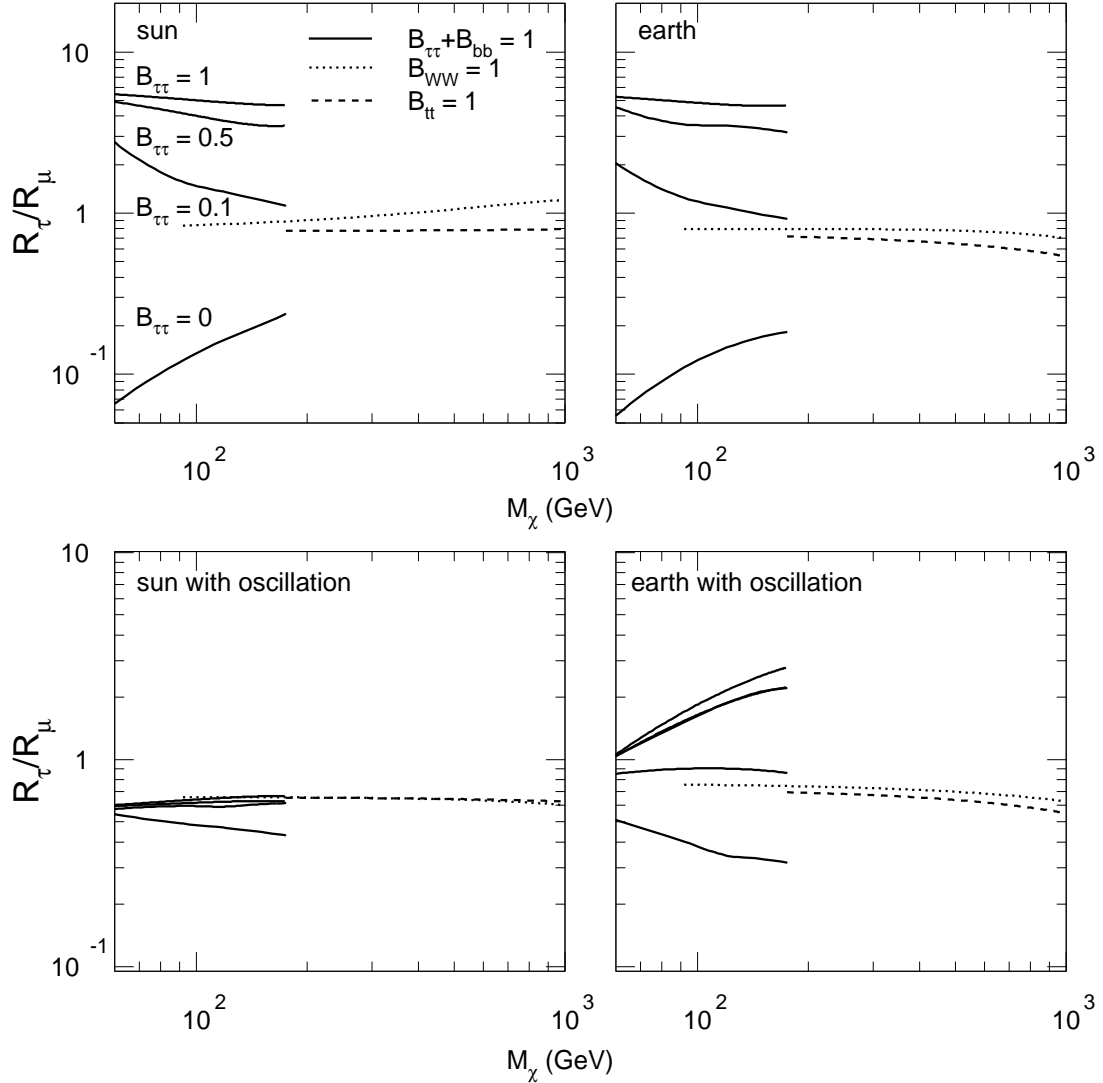}}
\caption{Ratios of CC contained tau-event rates, $R_{\tau}$, to  muon-neutrino event rates (contained and semi-contained), $R_{\mu}$, in the absence (upper part) and presence (lower part) of $\nu_{\tau}\leftrightarrow\nu_{\mu}$ oscillations. Annihilation into fermions ($B_{\tau\tau}+B_{bb}=1$) is represented by the four lines. The case of annihilation into W-bosons (top-quarks) is represented by the dotted (dashed) line.  
A detector of km $\times$ km $\times$ km  dimensions with an energy threshold of $E_{\nu}=25$~GeV has been assumed.}
\label{fig:rate_cont}
\end{center}
\end{figure}
Figure \ref{fig:rate} shows the effect of 
$\nu_{\tau}\leftrightarrow\nu_{\mu}$ oscillations on the neutrino-induced muon flux for the sun and the earth in the extreme cases discussed above, as well as for some moderate cases, motivated in section \ref{sec:MSSM}. 
The oscillation case is normalized to the no-oscillation case.  
When annihilating exclusively into a pair of taus, the muon flux increases due to oscillations up to a factor of 4-6 for the sun and 3-4 for the earth, which is explained by the higher and harder tau-neutrino flux~\cite{Seckel}. 
This enhancement is only weakly dependent on the exact value of the branching ratio, and even a small branching ratio of $B_{\tau\overline{\tau}}=0.1$ leads to considerably higher muon fluxes.
Only for a ratio of $B_{\tau\overline{\tau}}/B_{b\overline{b}}\lapproxeq 0.05$, a deficit due to neutrino oscillations can be expected. By scanning a large range of MSSM parameters as well as from equations (\ref{eq:higgs}) and (\ref{eq:bino}), we find that such a small ratio is below of what is expected in most MSSM models. 
Above the top-mass threshold,  annihilation into taus as final states can be neglected. Here,
the difference between the two extremes $B_{WW}$=1 and $B_{tt}$=1 is small. Due to the long baseline in the case of the sun, oscillations lead  to a small increase in the rate. For the earth, oscillations produce no visible effect, since for higher neutralino-masses the average oscillation length exceeds the radius of the earth considerably. Annihilation into $Z^0Z^0$ produces a rather similar ratio as annihilation into  $W^+W^-$.

Generally we find, that for a wide range of MSSM models, $\nu_{\tau}\leftrightarrow\nu_{\mu}$ oscillations lead to equal or larger muon fluxes. 
We note that this is a different conclusion than drawn in \cite{Fornengo}, where no increase but a generic decrease in rates due to neutrino-oscillations was found.

The ratio of contained tau-neutrino event rates, $R_{\tau}$, to the muon-neutrino event rate, $R_{\mu}$,
is shown in  figure \ref{fig:rate_cont} for the case of no oscillations (upper part) and the case of $\nu_{\tau}\leftrightarrow\nu_{\mu}$ oscillations (lower part). Here,  only detection by CC-reactions has been considered. The muon event rate includes contained and semi-contained events, where the vertex is located outside of the detection volume.
It was already shown in \cite{edsjo} that the contribution of the semi-contained events is small for a detector of the large size assumed here. With increasing WIMP mass, the average muon range grows and therefore leads to a decrease in the ratio $R_{\tau}/R_{\mu}$. This is  seen for the case of the earth, but for the sun, this effect is  compensated by the larger attenuation of the muon-neutrinos within the sun. 
If no oscillation is present, the event rate due to tau-neutrinos is significantly larger than due to muon-neutrinos, as long as the branching ratio $B_{\tau\tau}$ is large. 
In the case of $\nu_{\tau}\leftrightarrow\nu_{\mu}$ oscillations,  higher tau-neutrino event rates are only preserved for WIMP annihilation in the earth. For the sun, the flux of tau- and muon neutrinos completely average out, and the difference in event rates are due to the lower cross-section of the tau neutrino and the additional semi-contained muon events.

\section{Discussion}
After reviewing briefly some of the MSSM phenomenology relevant for 
tau-neutrino production, we have discussed the propagation of WIMP-neutrinos through the sun and the earth under consideration of the presently most favored $\nu_{\tau}\leftrightarrow\nu_{\mu}$ oscillation-parameters,  $\Delta m^{2}=0.003~$eV$^{2}$ and $\sin2\theta=1$~\cite{osci_par}. A main result is that in a first approximation,  scattering, absorption and regeneration of neutrinos in the sun can be effectively thought of as decoupled from oscillation-effects. Therefore the attenuation of the neutrino beam can be calculated first, and effects due to flavor-mixing can be applied later.

We have shown that, $\nu_{\tau}\leftrightarrow\nu_{\mu}$ oscillations can alter the muon flux from  
WIMP-annihilation significantly. In particular if the lightest neutralino is  B-ino like and lighter than the top-quark, we expect an increase of the 
muon flux of up to a factor of 4-6. If annihilation happens primarily through 
gauge-bosons or the top-quark, oscillations will lead to only small changes in 
the muon-flux, as the disappearing of the original muon-neutrinos,  $\nu_{\mu}\rightarrow\nu_{\tau}$,  is  
compensated by the appearance of original tau-neutrinos, $\nu_{\tau}\rightarrow\nu_{\mu}$. A decrease due to $\nu_{\tau}\leftrightarrow\nu_{\mu}$
oscillations is observed only if annihilation happens primarily into fermions 
and if $B_{\tau\tau}/B_{bb} \lapproxeq 0.05$, a ratio of branching ratios which is lower than expected 
for most of the MSSM parameter space.  

Because of the significantly higher tau-neutrino flux for parts of the MSSM 
parameter space, contained tau-events are a promising signal if a large 
detection volume is available.  However, the high tau-neutrino flux can be, 
relative to the muon-neutrino flux, considerably reduced if muon and 
tau-neutrinos mix. We omit electron-neutrinos from our discussion, since their flux at the origin is very similar to that of the muon-neutrinos. 

When comparing the detection potential for the different channels it is important to note that the background of atmospheric neutrinos sets limits to the WIMP-neutrino sensitivity, and  that the 
backgrounds due to  muon-, electron- or tau-neutrinos are significantly different in rate.  
In the relevant energy range, the flux of atmospheric muon-neutrinos is 10-20 times higher than the electron-neutrino flux \cite{Lipari} and  unless $\nu_{\mu}\rightarrow\nu_{\tau}$ oscillations are present, the flux of tau-neutrinos is negligible \cite{Stanev}. Additionally there is also a small neutrino-background from the sun itself \cite{Mannheim}. Experiment-related constraints might favor the detection of the CC muon-neutrino events which  generally consist of long muon-tracks in contrast to  tau- or electron-neutrinos producing cascade-like events. 
However, since contained cascade-like events deposit their energy within the detector we can expect a good energy resolution. This might add a new option to the rejection of atmospheric-neutrino events \cite{Bergstrom:1997tp}.      
In case of a measurable WIMP signal rate, an energy spectrum of cascade-like events could be used to get a rough estimate of the WIMP-mass. By comparing the fluxes of  muon- and tau-neutrinos it might become possible to obtain a hint on the composition of the WIMP. 
     
 
\section*{Acknowledgments}
This work was started during a research visit at the BRI with support of the Office of the Polar Programs of the U.S. National Science Foundation.
It is a pleasure to thank M.~Drees, J.~Edsj\"o, R.~Engel, M.~Lindner and D.~Seckel for useful discussions and comments as well as T.~Gaisser for the great hospitality received at the BRI.  The author would also like to thank C.~Spiering and C.~Wiebusch for careful reading of this manuscript. 


\end{document}